\documentclass[a4paper,twoside]{article}

\usepackage{epsfig}
\usepackage{subcaption}
\usepackage{calc}
\usepackage{amssymb}
\usepackage{amstext}
\usepackage{amsmath}
\usepackage{amsthm}
\usepackage{multicol}
\usepackage{pslatex}
\usepackage{apalike}
\usepackage{algorithm}
\usepackage{algpseudocode}
\usepackage{multirow}
\usepackage{url}

\usepackage{booktabs}
\usepackage[table]{xcolor}
\usepackage{graphicx}

\usepackage[bottom]{footmisc}
\usepackage[hidelinks]{hyperref}
\hypersetup{
    colorlinks=true,      
    linkcolor=blue,       
    citecolor=blue,       
    filecolor=blue,       
    urlcolor=blue         
}
\usepackage{SCITEPRESS}     

\begin{document}

\title{PromptMark: A Prompt-Guided Iterative-Feedback Framework for Source Code Watermarking}

\author{
    \authorname{
        Istiaq Ahmed Fahad,
        Mridha Md. Nafis Fuad,
        Kazi Sakib
    }
    \affiliation{
        Institute of Information Technology, University of Dhaka, Bangladesh
    }
    \email{
        \{bsse1204, fuad, sakib\}@iit.du.ac.bd
    }
}


\abstract{
Watermarking has become a crucial technique for ensuring provenance and accountability in AI-generated source code. As large language models (LLMs) are increasingly integrated into development workflows, reliable attribution remains challenging. In practice, most developers rely on commercial LLM APIs operating under black-box constraints, making existing approaches that require access to the decoding process less feasible for real-world integration. To address this limitation, we propose PromptMark, a black-box, prompt-guided watermarking framework that embeds invisible yet statistically detectable signals into generated code via structured input instructions. The method steers models toward subtle identifier and comment naming patterns while preserving the functional correctness and structural integrity of the generated code. Detection is performed using statistical tests designed to remain reliable across varying code lengths and model outputs. The embedding is further refined through an iterative feedback loop, where prompts are updated based on watermark detection scores. Experiments on the MBPP and HumanEval benchmarks show that PromptMark consistently achieves strong watermark detectability while maintaining high code correctness, outperforming baseline approaches.
}

\onecolumn \maketitle \normalsize \setcounter{footnote}{0} \vfill

\section{\uppercase{Introduction}}
\label{sec:introduction}

Watermarking refers to the process of embedding a hidden but verifiable signal into digital content, enabling author attribution or provenance tracking. It has long been recognized as an effective technique for protecting the ownership and integrity of digital artifacts ~\cite{collberg_thomborson_2002}. While traditional watermarking methods have been widely applied to images, audio, and text ~\cite{hayes2017generating,liu2024audiomarkbench}, extending this concept to source code introduces distinct challenges. Source code is highly structured and syntactically rigid, where even minor modifications can break compilation or alter semantics. At the same time, Large Language Models (LLMs) are being widely used by developers to generate and modify code, making reliable attribution of AI-generated programs an important concern. Despite this concern, designing an effective watermarking approach while preserving code correctness remains a challenging task. 

In practical development settings, most users interact with LLMs through commercial API-based services that operate under strict black-box constraints, where only input and output are accessible ~\cite{bistarelli_fiore_mercanti_mongiello_2025}. This limitation restricts direct control over the generation process, making it difficult to embed reliable watermark signals without compromising code correctness. Existing watermarking approaches are largely designed under white-box assumptions, where access to model internals enables manipulation of token probabilities or decoding strategies ~\cite{lee2024,guan2024codeip}. While effective in controlled environments, their applicability in practical deployment settings remains limited.

Under these constraints, watermarking methods should be designed that do not require access to model internals. Prompt-based methods provide a practical alternative, as they utilize input instructions to influence generation behavior without modifying the underlying model ~\cite{dasgupta2024watermarking}. In particular, they exploit the instruction following capabilities of modern LLMs to steer outputs toward desired patterns. However, effectively embedding watermark signals through prompts remains challenging, as it requires balancing the inclusion of detectable patterns with the preservation of syntactic correctness and functional behavior of the generated code.

Several studies have explored different watermarking techniques for AI-generated code, such as influencing token selection through entropy-thresholding ~\cite{lee2024}, enforcing grammar-guided constraints to support multi-bit embedding ~\cite{guan2024codeip}, or biasing token prediction by manipulating non-syntax tokens ~\cite{kim2025marking}. Recent studies also explore attention-based triggers to determine watermarking location for improved robustness ~\cite{li2025codeguard}. Although effective, their applicability in black-box settings is limited due to reliance on model internals. 

To address these limitations, we propose \textbf{PromptMark}, a prompt-guided black-box watermarking framework for source code. This framework operates in three phases. In the first phase, PromptMark constructs a system prompt that guides the model to generate identifiers and comments whose initial characters follow a predefined adaptive green list, which serves as the watermark signal. This list is derived through frequency-based stratification of identifier initials observed in human-written code. In the second phase, the model generates code following an iterative feedback loop that checks for both functional correctness and sufficient watermark strength. When either condition is not satisfied, a summarized feedback log is incorporated into the subsequent prompt to refine the output. Finally, the detection phase identifies watermarks by extracting the initial characters from identifiers and comments in the generated code. A p-value is then computed based on the observed frequency of green list characters to quantify the likelihood of watermark presence.

We evaluated PromptMark on Python code using two closed-source models: Gemini 2.5 Flash and Claude Sonnet 4 on MBPP and HumanEval datasets. We focus on two core aspects: effectiveness and code quality of the generated watermarked code. Across different configurations, the iterative-feedback method with an adaptive green list achieves strong detection performance, with AUROC  ranging from approximately 0.90 to 0.98. At a low false positive rate (1\%--5\% FPR), the method showed high true positive rates, exceeding 90\% in the majority of configurations and reaching above 95\% in favorable settings. In terms of code quality, the framework preserves functional correctness in around 90\% of the cases, while maintaining nearly similar CodeBLEU scores to non-watermarked code.

\section{\uppercase{Problem Definition AND Related Work}}
This section formalizes the problem of watermarking and reviews existing studies for establishing authorship and origin of software artifacts. We first introduce the concept of code provenance, followed by two primary paradigms: active source code watermarking and passive authorship attribution. We then discuss existing techniques within these paradigms and highlight their limitations in practical deployment settings.

\subsection{The Concept of Code Provenance}
Code provenance is defined as the verifiable, auditable record of the origin, authorship, and modification history of software. It functions as a ``chain of custody'' for code, ensuring that every artifact in production can be traced to specific, authorized commits and builders. In environments where AI and humans now work as co-authors, the origin of code cannot be inherently trusted. Rather, the identity of the coder or maintainer becomes the anchor of trust. Accordingly, every commit, whether written by a human or generated by an AI, must be signed and bound to a verified identity to ensure accountability rather than restricting automation. This principle is formalized in frameworks such as Supply-chain Levels for Software Artifacts (SLSA), which, under policies influenced by Executive Order 14028 \cite{murray_2025}, requires verifiable proof of authorship and build integrity to sell software to the U.S. government ~\cite{gu_2025}. 

Previous studies ~\cite{xiang_review_25} address provenance from two different angles: watermarking ~\cite{guan2024codeip} and authorship attribution~\cite{gurioli2025you}.

\subsection{Watermarking}
Watermarking is a way to hide a secret signal inside a carrier, such as text or images. It is used to prove ownership or check whether the content is authentic ~\cite{collberg_thomborson_2002}. When we apply this to a generative model, it involves changing or biasing the model's token prediction process to leave a detectable mark in the generated output.

Source code watermarking, on the other hand, is more challenging than natural language because we must additionally ensure that the watermarked version of the code is functional and executable~\cite{guan2024codeip}. Unlike a story or an essay, code is fragile: a single wrong character can break functionality or even alter the program logic for which the code was initially written. ~\cite{lee2024}. Therefore, a code watermark must modify the code in such a way that the watermarking signal is hidden and the logic and functionality remain intact.

Formally, a watermarking system for a model $\mathcal{M}$ consists of two parts $(\mathcal{G}, \mathcal{D})$ ~\cite{srcmarker}:

\begin{enumerate}
    \item \textbf{Generation ($\mathcal{G}$):} The system takes a prompt $x$ and a secret key $k$ to produce watermarked code $y_w$ following the Equation ~\ref{eq:generation}:
    \begin{equation}
    \label{eq:generation}
         \mathcal{G}(x, \mathcal{M}, k) \rightarrow y_w 
    \end{equation}
During this step, the model is biased to favor certain tokens determined by the key $k$ ~\cite{kirchenbauer2023watermark,lee2024}. 

\item \textbf{Detection ($\mathcal{D}$):} The system takes a suspect code snippet $\hat{y}$ and the secret key $k$ to decide whether the code is watermarked. The detector computes a continuous statistical score representing the strength of the embedded signal, defining the detection as a formal hypothesis testing problem:
    \begin{itemize}
        \item $H_0$: The code $\hat{y}$ is generated without the knowledge of $k$ (not watermarked).
        \item $H_1$: The code $\hat{y}$ is generated with the knowledge of $k$ (watermarked).
    \end{itemize}
The detection function provides a binary decision by comparing this computed score against a predefined threshold $\eta$:
    \begin{equation}
    \label{eq:detection}
        \begin{aligned}
            \mathcal{D}(\hat{y}, k) &\rightarrow d , \quad d \in \{\text{true, false}\}.
        \end{aligned}
    \end{equation}
\end{enumerate}

The detector returns \textit{true} (rejecting $H_0$) when the score is $>\eta$, indicating that the code is ``watermarked.'' It returns \textit{false}, suggesting the observed pattern is likely due to random choice. As the detection process strictly verifies the binary presence or absence of a signal, it is formally categorized as \textit{zero-bit} watermarking ~\cite{zhao2025sok}. While \textit{multi-bit} schemes extend this concept by embedding recoverable payloads, our research here focuses exclusively on the \textbf{zero-bit} approach.

\subsection{Authorship Attribution}
Authorship attribution is a forensic method used to figure out who wrote a document based on how it was written. This process relies on ``Code Stylometry'', where the unique style of writing code provides a trace of the author. These stylistic traces are the fingerprints of the programmers that separate one's work from others ~\cite{gurioli2025you}. Researchers look for specific style features in the code to find these fingerprints. This includes lexical features like how variables are named or how comments are written ~\cite{park2025detecting}. It also includes structural features. These can be patterns like how deep loops are nested or the specific way the code logic is organized ~\cite{paek2025detection}. Some modern systems even use neural networks to learn these patterns automatically ~\cite{yin2025detecting,xu2024distinguishing}.

\subsection{{Related Work}}
\label{sec:related_works}
A series of studies has investigated in-process watermarking for code generation, where watermark signals are embedded during the decoding process. Early efforts include SWEET~\cite{lee2024}, which applies entropy-based logit adjustments to bias token selection toward statistically identifiable patterns. CODEIP~\cite{guan2024codeip} extends this idea by leveraging programming-language grammar to guide token prediction and support multi-bit watermark embedding. To improve naturalness and functional integrity, STONE~\cite{kim2025marking} limits watermarking to non-syntax tokens such as identifiers and literals. Beyond lexical manipulation, CodeGuard~\cite{li2025codeguard} introduces an attention-based approach that activates watermarking through learned triggers and harmonic character patterns. MCGMark~\cite{ning2024mcgmark} proposes an encodable watermarking approach for tracing malicious code. They manipulate vocabulary selection and skip specific tokens to enhance robustness. RoSeMery~\cite{zhang2025robust} adopts an ML/crypto co-design, embedding multi-bit signals through encoder-guided renaming and structural modifications while safeguarding watermark keys. Although diverse in technique, all of these methods rely on access to model internals, which makes them limited to white-box approaches.

Separately, a growing body of research examines prompt-level watermarking, where watermark signals are embedded solely through natural-language instructions. In-context watermarking (ICW)~\cite{liu2025} demonstrates that lexical preferences can be imposed through system prompts, while prompt-guided multi-LM frameworks~\cite{dasgupta2024watermarking} show how watermark instructions, generation, and detection strategies can all be produced through coordinated prompting. However, these studies focus primarily on natural-language generation rather than source code. CodeMark-LLM~\cite{xu2025large} adopts a post-hoc, multi-bit watermarking strategy that uses LLMs to embed watermarks by refactoring existing code via pre-defined transformation rules. This approach requires access to the original code for watermark verification. ACW~\cite{li2024acw} employs a training-free, zero-bit detection framework that embeds watermarks via idempotent, semantic-preserving transformations.

Despite these advancements, prompt-based watermarking for code-generation tasks in black-box, API-only settings remains largely unexplored. To the best of our knowledge, no prior work has investigated whether carefully crafted prompts can reliably embed detectable watermark signals in source code.

\begin{table}[htbp]
    \centering
    \caption{Identifier types and descriptions}
    \label{tab:variables}
    \begin{tabular}{p{0.33\columnwidth} p{0.57\columnwidth}}
        \toprule
        \textbf{Identifier Type} & \textbf{Description} \\
        \midrule
        Function Names & User-defined function names \\
        Class Names & User-defined class names \\
        Variable Names & Local/global variable identifiers \\
        Function Arguments & Parameter names in function definitions \\
        Attribute Names & Object attributes or method references \\
        Loop Variables & Iterator variables in loop constructs \\
        \bottomrule
    \end{tabular}
\end{table}

\begin{figure}
    \centering
    \includegraphics[width=0.99\linewidth]{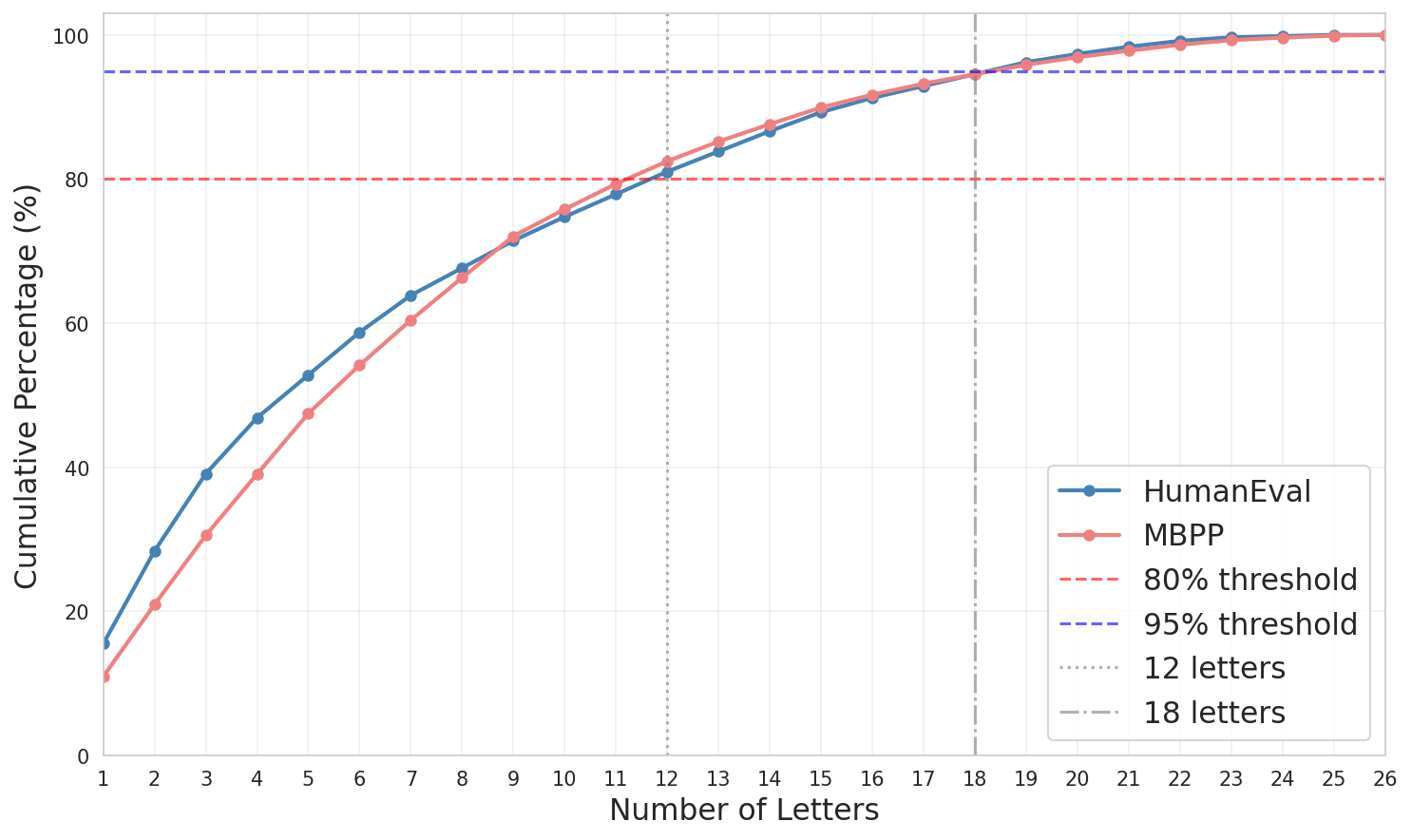}
    \caption{\centering{Combined cumulative initial letter distributions (HumanEval / MBPP)}}
    \label{fig:letter-pct}
\end{figure}

\begin{figure*}[htbp]
    \centering
    \includegraphics[width=0.96\textwidth]{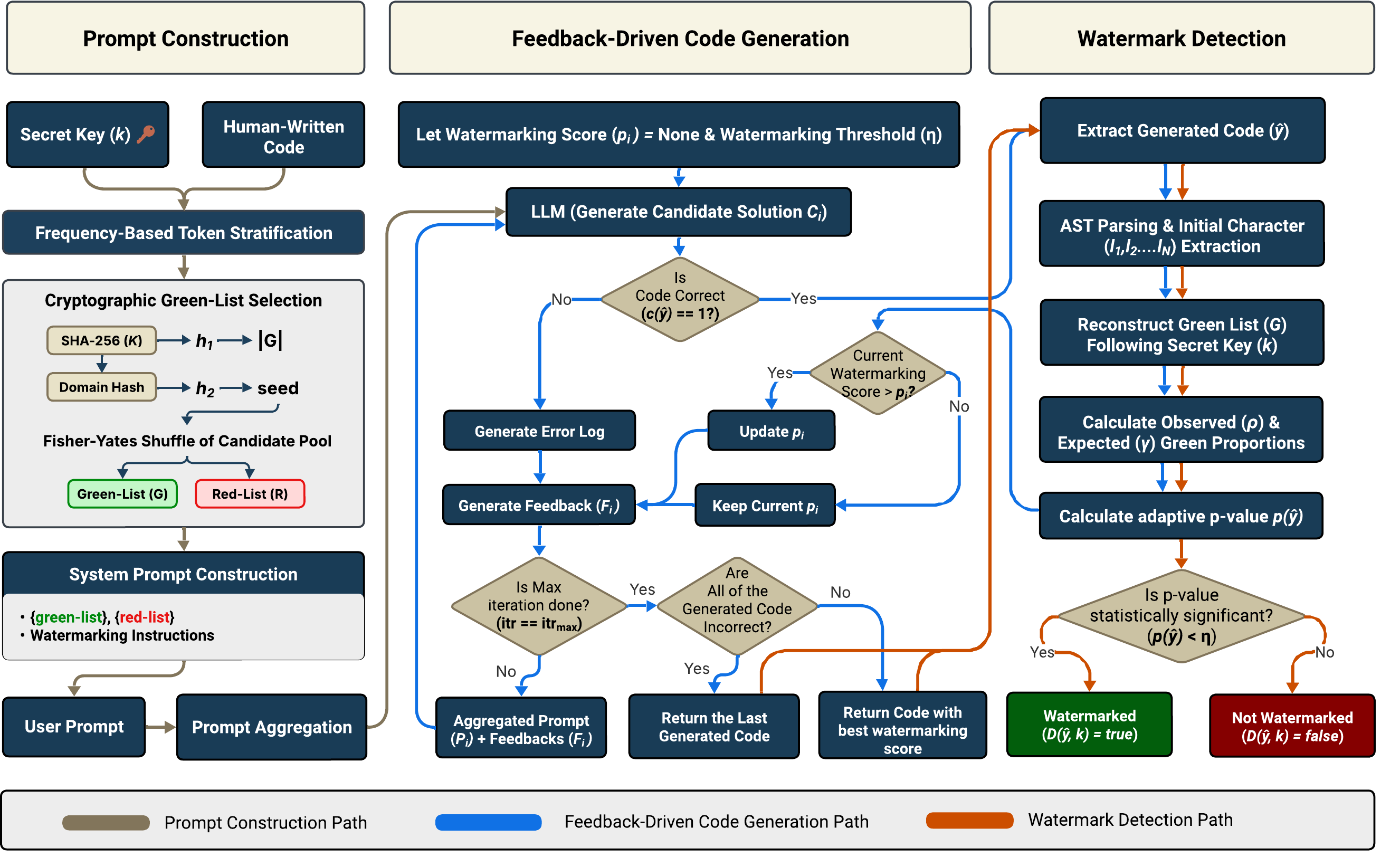}
    \caption{Overview of the PromptMark framework.}
    \label{fig:methodology}
\end{figure*}

\section{\uppercase{Methodology}}
\label{sec:methodology}

We propose PromptMark, a prompt-guided watermarking framework for source code that embeds imperceptible yet statistically detectable watermarks into LLM-generated code. Prior studies ~\cite{lee2024,kim2025marking} assume white-box access and embed signals by directly biasing token logits during decoding. In contrast, we target closed-source models such as Gemini and Claude, where logit manipulation and decoder modification are unavailable. Moreover, large-scale parameter updates or fine-tuning are often restricted or economically impractical. Consequently, under black-box constraints, watermarking must rely on input-level control, such as prompt instructions to steer the generation process, which motivates our approach. Our three-step method embeds watermarks without accessing model internals, instead leveraging LLMs' instruction following capabilities.

\subsection{Prompt Construction}
In this step, we construct a system prompt that biases the model toward producing watermark-carrying tokens specifically, identifiers and comments whose \emph{initial characters} belong to a predefined \emph{green list}. Building on the general ``green list/red list'' watermarking paradigm introduced in WLLM~\cite{kirchenbauer2023watermark}, we define the \emph{green list} as the subset of allowed initial characters that encodes the watermark signal. The complementary \emph{red-set} contains characters that the model is instructed to avoid as initial characters for the targeted tokens. Enforcing this constraint enables a statistically detectable skew in initial-character frequencies that distinguishes watermarked outputs from non-watermarked ones.

However, identifier names in source code cannot be modified arbitrarily as they might break dependencies, public interfaces, or library conventions. Therefore, we restrict watermarking to a set of ``safe'' Python identifier categories (see Table~\ref{tab:variables}) that can be altered without affecting functionality. To generate the green and red sets, we perform frequency-based stratification over the initial characters of identifiers observed in human-written code  ~\cite{lee2024}. Finally, to increase embedding capacity beyond identifiers, we additionally watermark comments using the same initial-character constraint. The complete prompt is presented in Figure~\ref{fig:system-prompt}.

\begin{figure}[htb]
\centering
\fbox{%
\begin{minipage}{0.96\linewidth}
\raggedright
\setlength{\emergencystretch}{3em}
\scriptsize

You are an expert Python programmer and code reviewer. Your task is to write a function that adheres to specific identifier naming rules and generates code based on a given problem statement.

\textbf{\#\# Additional Instruction:}

\textbf{\#\#\# Green Letter List:} \{green\_words\}

\textbf{\#\#\# Red Letter List:} \{red\_words\}

\vspace{4pt}
\textbf{\#\#\# Command:}

Generate code following the given instructions:

\textbf{1. Green Letter Enriched Identifier:}

When generating identifiers (e.g., local variables, parameters, helper functions, attributes, temporary variables) and comments, start words with letters from the \textbf{Green Letter List}. Use examples as naming references.

\vspace{2pt}
\textbf{2. Correct \& Relevant:}

Generate correct code following the problem statement.

\vspace{2pt}
\textbf{3. About comments:}

Add brief comments only for complex logic. Avoid tokens starting with \textbf{Red List} letters.

\vspace{2pt}
\textbf{4. About Method Name:}

Use the exact method name from the given test case.

\vspace{2pt}
\textbf{5. Warning:}

Do not mention or explain renaming rules in output.

\vspace{2pt}
\textbf{6. Others:}

Explain exactly three bullet points after the code. If incorrect, add corrected code after explanation. Never include test cases, explanations, or assertions inside code blocks.

\end{minipage}%
}
\caption{System prompt used to constrain and bias the code generation process.}
\label{fig:system-prompt}
\end{figure}

\subsection{Frequency-Based Token Stratification}
\label{sub-freq-stratification}

To embed a watermark without introducing unnatural or forced identifier names, we analyze the distribution of identifier-initial letters in human-written Python code using the MBPP \cite{austin2021programmbpp} and HumanEval \cite{chen2021evaluatinghumaneval} datasets. This allows the watermark to bias initial characters while preserving semantically meaningful and human-like naming conventions. The observed rank-frequency distribution ($f$) of the character initials follows a Zipf-like pattern~\cite{piantadosi2014zipf}, where a relatively small subset of letters (12-18) accounts for a large proportion of identifier initials (see Figure~\ref{fig:letter-pct}).

Since both MBPP and HumanEval primarily contain short-to-medium length programs with a limited number of unique identifiers, the expected natural occurrence rate $\gamma$ must remain moderate to ensure statistical separability between watermarked and non-watermarked code. If $\gamma$ is too high, most identifiers would naturally satisfy the green list constraint, reducing discriminative power in the hypothesis test. To determine an appropriate candidate-pool size, we compute the empirical baseline probability $\gamma$ for varying values of $K$ (number of top frequent letters). 

To determine the suitable range, we conducted experiments over a broader spectrum of $K$ values. The results consistently exhibit a monotonic increasing trend in $\gamma$ with respect to $K$, confirming that larger candidate pools lead to higher baseline probabilities and reduced statistical separability. Thus, given the moderate code length in the MBPP and HumanEval datasets, we select a candidate-pool range that keeps $\gamma$ within a balanced interval, preserving both naturalness and statistical detectability. Accordingly, we define the candidate set $C$ as the top-$K$ most frequent initial characters:
\begin{equation}
C = \operatorname{top}_K(f), \quad K \in [10,12].
\end{equation}
This range maintains a moderate baseline green-letter probability ($\gamma \approx 0.52$--$0.62$), allowing reliable segregation between watermarked and non-watermarked code while maintaining consistency with natural identifier distributions. While other values of $K$ follow similar trends, this interval provides the best trade-off between embedding strength and false positive control in our setting. However, in practical deployments, the baseline probability $\gamma$ should ideally be estimated from the target developer’s codebase to reflect their natural identifier usage patterns.

\subsubsection{Cryptographic Derivation of the Green List}

To avoid fixed cardinality and predictable identifiers, we derive the green list size $|G|$ and its specific letter composition directly from a 128-bit secret key $k$. The complete deterministic extraction procedure is summarized in Algorithm~\ref{alg:green_list}. 

\textbf{Uniform Sampling of Candidate-Pool Size.}
We utilize the SHA-256 hash function to approximate uniform randomness (Line 1). By extracting the first 32 bits of the hashed key, we obtain a uniformly distributed continuous variable in $[0,1)$ (Line 2). This variable is linearly scaled to uniformly sample the cardinality $|G|$ from the admissible bounds $[|G|_{\min}, |G|_{\max}]$ (Line 3). This ensures that every permissible size for the green list has an equal probability of being selected, preventing predictable patterns in the watermark's structure.

\textbf{Domain-Separated Unbiased Permutation.}
To determine which specific letters form $G$, we compute a second, domain-separated hash (Line 4) to generate a 64-bit integer seed (Line 5). This seed drives a Fisher-Yates shuffle~\cite{knuth1997art} over the empirically grounded candidate pool $C$ (Line 6). The Fisher-Yates algorithm guarantees an unbiased permutation, ensuring every possible ordering of $C$ is produced with equal probability. The first $|G|$ letters of this permuted list form the green list $G$, while the remaining letters form the complement red-set $R$ (Lines 7--8).

\begin{algorithm}[htbp]
\caption{Cryptographic Green List Construction}
\label{alg:green_list}
\begin{algorithmic}[1]
\renewcommand{\algorithmicrequire}{\textbf{Input:}}
\renewcommand{\algorithmicensure}{\textbf{Output:}}

\Require Secret key $k$, Candidate pool $C$, Cardinality bounds $|G|_{\min}, |G|_{\max}$
\Ensure Green List $G$, Red List $R$
\vspace{0.15cm}

\State $h_1 \gets \text{SHA-256}(k)$
\State $u_1 \gets \text{bytes\_to\_int}(h_1[0:4]) / 2^{32}$ 
\State $|G| \gets \left\lfloor |G|_{\min} + u_1 \cdot (|G|_{\max} - |G|_{\min} + 1) \right\rfloor$
\vspace{0.15cm}

\State $h_2 \gets \text{SHA-256}(k \parallel \text{``green-set-selection''})$
\State $\text{seed} \gets \text{bytes\_to\_int}(h_2[0:8])$ 
\State $\texttt{p} \gets \text{Fisher-Yates-Shuffle}(C, \text{seed})$ 
\vspace{0.15cm}

\State $G \gets \{\texttt{p}[0], \texttt{p}[1], \dots, \texttt{p}[|G|-1]\}$
\State $R \gets C \setminus G$
\vspace{0.15cm}

\State \Return $G, R$
\end{algorithmic}
\end{algorithm}

The proposed green list generation strategy ensures reproducibility, security, and naturalness. Given the same secret key $k$, the same green list $G$ is deterministically generated, enabling consistent watermark embedding and verification in black-box settings. The use of the non-invertible SHA-256 hash function makes the mapping from $k$ to $G$ cryptographically secure, as recovering the 128-bit seed would require approximately $2^{128}$ operations. Besides, selecting frequently used identifier initials as green list characters preserves a semantically meaningful tone while creating a detectable watermark signal.


\subsection{Iterative Code Generation}
The watermark embedding process is designed as an iterative code generation process followed by feedback that jointly focuses on two primary objectives: \textbf{functional correctness} and \textbf{watermark fidelity} (see Figure~\ref{fig:methodology}). Here, watermark fidelity denotes how strongly and consistently the generated code adheres to the intended watermarking pattern without compromising correctness. When the constructed prompt is passed to the LLM, it generates candidate code following the watermarking instructions provided via the system prompt. Formally, at iteration $i \in [1, Itr_{\text{max}}]$, the model generates a candidate solution 
$C_i$ guided by prompt $P_i$. The evaluation of each candidate is defined as:
\begin{equation}
\label{eq:evaluate}
\text{Eval}(C_i) = (c_i,\, p_i),
\end{equation}
where $c_i$ is the correctness score based on unit-test execution ($c_i=1$ if all tests pass, otherwise $0$), and $p_i$ is the watermark fidelity score, which is a p-value score as calculated in the detection phase, Section ~\ref{subsec-wm-detection}.

After each generation, the LLM produces feedback $F_i$ to improve any unmet evaluation criteria. If $C_i$ fails the functional tests, the summarized error is returned as feedback. If the code is correct but the watermark score is below the threshold, the feedback instructs the model to use more identifiers starting with green list characters. The next prompt is formed as $P_{i+1}=f(P_i, F_i)$, where $f(\cdot)$ denotes the prompt-update operation that appends the feedback into the next prompt. This loop continues until both objectives are met or the iteration limit $Itr_{\max}$ is reached, forming the stopping criterion of iteration.


If no candidate satisfies both constraints, we return the candidate with the best watermark fidelity score $p_i$, representing the closest achievable watermark signal. This preserves the strongest available watermark signal instead of returning an arbitrary result. Thus, the iterative-feedback method steers the model toward candidate solutions that jointly improve watermark strength while preserving the correctness of the generated code.

\subsection{Watermark Detection}
\label{subsec-wm-detection}
The watermark detection is treated as a statistical test that checks whether a generated code shows a noticeable shift in the distribution of identifier initials compared to human-written code. This process is model-agnostic and works in a black-box setting without access to model parameters. 
Let $\hat{y}$ denote the final code produced by the embedding process. The code is parsed into an Abstract Syntax Tree (AST) to extract all identifiers. From each identifier $t_i$, we obtain its first alphabetic letter $l_i$, normalized to lowercase. Let $N$ be the total number of extracted initials, forming the sequence $L = [l_1, l_2, \dots, l_N]$.

Let $I$ denote the number of initials belonging to the green list $G$. The observed proportion is $\rho = I/N$, while the expected proportion $\gamma$ is estimated from human-written code. Watermark detection is framed as a one-sided hypothesis test as:
\[
H_0: \rho <= \gamma \quad \text{,} \quad H_1: \rho > \gamma,
\]
Here, $H_1$ indicates a statistically higher occurrence of green-initial identifiers. To ensure reliable detection across varying code lengths, we compute an adaptive 
\textbf{p-value} defined as
\begin{equation}
\label{eq:adaptive-score}
p =
\begin{cases}
\displaystyle\sum_{i=I}^{N} \binom{N}{i}\,\gamma^i (1-\gamma)^{N-i}, & \text{if } N < 30, \\[6pt]
1 - \Phi\!\left(\dfrac{I - N\gamma}{\sqrt{N\gamma(1-\gamma)}}\right), & \text{otherwise},
\end{cases}
\end{equation}

where $\Phi(\cdot)$ denotes the cumulative distribution function of the standard normal distribution. The p-value estimates the probability of observing at least $I$ green-initial identifiers out of $N$ under the null rate $\gamma$. We refer to it as an ``adaptive'' p-value because it switches between the exact binomial test ($N < 30$) and the normal approximation ($N \ge 30$) to ensure accurate and stable estimation, as commonly done in statistical detection and anomaly analysis~\cite{li2011application}.

For evaluation, we define a monotonically increasing ranking statistic
$S = -\log_{10}(p)$, where larger $S$ indicates stronger evidence of watermarks. This transformation improves dynamic range for small $p$-values and is used to compute the AUROC score.

To conduct the hypothesis test, we directly compare the 
adaptive p-value $p(\hat{y})$ against a predefined significance 
threshold $\eta$. Let $c(\hat{y}) \in \{0,1\}$ denote the functional 
correctness indicator, where $c(\hat{y})=1$ if the generated code passes 
all unit tests and $0$ otherwise. Having the secret key as $k$, the final detection decision, illustrated in Figure~\ref{fig:methodology}, is defined as:
\begin{equation}
\label{eq:detection}
D(\hat{y},k)=
\begin{cases}
\text{true}, & \text{if } c(\hat{y}) = 1 \;\land\; p(\hat{y}) < \eta, \\
\text{false}, & \text{otherwise}.
\end{cases}
\end{equation}

Thus, following Equation \ref{eq:detection}, a code is classified as watermarked only if it is functionally correct and its p-value falls below the significance threshold $\eta$.

\section{\uppercase{Results and Evaluation}}
\label{sec:results}
We evaluate PromptMark across multiple watermarking strategies and two benchmark datasets, MBPP and HumanEval, using Gemini 2.5 Flash and Claude Sonnet 4. All model outputs were generated using the providers' default hyperparameters (temperature=1.0, top-p=0.95, top-k=64). The baseline methods were also evaluated under the default settings provided in their work. The following are the research questions we investigated through our experiments:

\begin{itemize}
\item \textbf{RQ1 (Effectiveness):} How effective is PromptMark in embedding and detecting source code watermarks?
\item \textbf{RQ2 (Code Quality):} What impact does PromptMark have on code quality?
\end{itemize}


\begin{table*}[htbp]
\centering
\caption{\centering{Performance comparison of watermarking methods across MBPP and HumanEval datasets (↑ higher or ↓ lower indicates the best values for each metric). Best results are in \textbf{bold}; grayed rows denote our final methods}}
\label{tab:results}
\resizebox{\textwidth}{!}{%
\begin{tabular}{ll cccccc cccccc}
\toprule
\multirow{2}{*}{\textbf{Models}} & \multirow{2}{*}{\textbf{Method}} & \multicolumn{6}{c}{\textbf{MBPP}} & \multicolumn{6}{c}{\textbf{HumanEval}} \\
\cmidrule(lr){3-8} \cmidrule(lr){9-14}
& & \textbf{Pass\% ↑} & \textbf{Avg Itr. ↓} & \textbf{AUROC ↑} & \textbf{T@1\%F ↑} & \textbf{T@5\%F ↑} & \textbf{CodeBLEU ↑} 
& \textbf{Pass\% ↑} & \textbf{Avg Itr. ↓} & \textbf{AUROC ↑} & \textbf{T@1\%F ↑} & \textbf{T@5\%F ↑} & \textbf{CodeBLEU ↑} \\
\midrule

Qwen & SWEET & 43.83 & - & 0.920 & 0.440 & 0.720 & 0.420 & 26.95 & - & 0.810 & 0.450 & 0.540 & 0.310 \\
\midrule

& ICW & 94.44 & - & 0.512 & 0.045 & 0.072 & 0.440 & 86.70 & - & 0.400 & 0.010 & 0.040 & 0.480 \\
& No Watermark & 64.77 & - & 0.577 & 0.035 & 0.106 & 0.420 & 85.50 & - & 0.530 & 0.040 & 0.090 & 0.480 \\
& Static & 98.25 & - & 0.922 & 0.398 & 0.597 & 0.440 & 90.00 & - & 0.810 & 0.160 & 0.350 & 0.490 \\
& Refactoring-Based & 99.45 & - & 0.879 & 0.111 & 0.422 & \textbf{0.910} & 95.63 & - & 0.890 & 0.210 & 0.510 & \textbf{0.500} \\
& Comment-based & 95.01 & - & 0.925 & 0.418 & 0.686 & 0.440 & 88.90 & - & 0.850 & 0.330 & 0.510 & 0.470 \\

\rowcolor{gray!20}
& Iterative & \textbf{99.78} & \textbf{1.20} & 0.974 & 0.707 & 0.891 & 0.460 
& \textbf{97.86} & \textbf{1.90} & 0.950 & 0.750 & 0.830 & 0.480 \\

\rowcolor{gray!20}
\multirow{-7}{*}{Gemini} 
& Iterative-AGL & 96.79 & 1.90 & \textbf{0.994} & \textbf{0.967} & \textbf{0.979} & 0.460 
& 88.90 & 1.93 & \textbf{0.970} & \textbf{0.900} & \textbf{0.900} & 0.480 \\

\midrule

& ICW & 92.40 & - & 0.770 & 0.020 & 0.140 & 0.500 & 94.95 & - & 0.740 & 0.010 & 0.120 & 0.490 \\
& No Watermark & 63.96 & - & 0.416 & 0.040 & 0.060 & 0.500 & 96.28 & - & 0.362 & 0.018 & 0.030 & 0.490 \\
& Static & 86.73 & - & 0.899 & 0.260 & 0.560 & 0.460 & 95.64 & - & 0.773 & 0.240 & 0.410 & 0.490 \\
& Refactoring-Based & \textbf{99.67} & - & 0.723 & 0.080 & 0.080 & \textbf{0.650} & \textbf{97.86} & - & 0.693 & 0.013 & 0.15 & \textbf{0.790} \\
& Comment-based & 90.00 & - & 0.892 & 0.390 & 0.600 & 0.470 & 91.86 & - & 0.855 & 0.420 & 0.590 & 0.490 \\

\rowcolor{gray!20}
& Iterative & 97.27 & \textbf{1.33} & \textbf{0.989} & 0.900 & 0.960 & 0.460 
& 97.32 & \textbf{1.31} & 0.903 & 0.563 & 0.680 & 0.500 \\

\rowcolor{gray!20}
\multirow{-7}{*}{Claude} 
& Iterative-AGL & 93.60 & 2.50 & 0.987 & \textbf{0.870} & \textbf{0.940} & 0.470 
& 94.00 & 1.83 & \textbf{0.930} & \textbf{0.640} & \textbf{0.710} & 0.480 \\

\bottomrule
\end{tabular}}
\end{table*}

\subsection{Evaluation Metrics}

We evaluate PromptMark following the established metrics as used in previous studies~\cite{lee2024,kim2025marking}. These include evaluation on functional correctness (Pass\%), watermark detectability (AUROC, TPR, FPR, \text{T@x\%F}), and code quality (CodeBLEU) metrics.

\paragraph{Pass\% (Functional Correctness).}
Pass\% measures the percentage of generated programs that pass all unit tests:
\begin{equation}
\text{Pass\%} = \frac{\text{\# Correct Solution}}{\text{Total Problems}} \times 100
\end{equation}
Here, this metric reflects whether watermark embedding preserves program functionality.

\paragraph{True Positive Rate (TPR).}
TPR measures the proportion of watermarked samples correctly identified:
\begin{equation}
\text{TPR} = \frac{\text{TP}}{\text{TP} + \text{FN}},
\end{equation}
where TP denotes watermarked code correctly detected as watermarked, and FN denotes watermarked code incorrectly classified as non-watermarked.  
In our scenario, a higher TPR means the detector successfully recognizes embedded watermarks.

\paragraph{False Positive Rate (FPR).}
FPR measures the proportion of non-watermarked samples incorrectly classified as watermarked:
\begin{equation}
\text{FPR} = \frac{\text{FP}}{\text{FP} + \text{TN}},
\end{equation}
where FP denotes clean (non-watermarked) code falsely detected as watermarked, and TN denotes correctly rejected clean samples.  
A low FPR is critical to avoid falsely attributing watermark ownership to benign code. 
We also measure detection performance under strict false-positive rates, such as T@1\%F and T@5\%F, where $\text{T@x\%F} = \text{TPR} \;\text{subject to}\; \text{FPR} \leq x\%$


\paragraph{AUROC (Area Under the ROC Curve).}
AUROC measures the overall separability between watermarked and non-watermarked samples across all decision thresholds:
\begin{equation}
\text{AUROC} = \int_{0}^{1} \text{TPR}(t)\, d(\text{FPR}(t)).
\end{equation}
An AUROC of 0.5 indicates random guessing, while values close to 1.0 indicate strong discriminative capability.

\paragraph{CodeBLEU (Code Quality).}
CodeBLEU evaluates structural and semantic similarity between generated and reference code:
\begin{equation}
\begin{aligned}
\text{CodeBLEU} &=
\alpha \cdot \text{Ngram}
+ \beta \cdot \text{WeightedNgram} \\
&\quad + \gamma \cdot \text{AST}
+ \delta \cdot \text{Dataflow},
\end{aligned}
\end{equation}
where $\alpha, \beta, \gamma, \delta$ are weighting coefficients.

Recent findings by ~\cite{mahmud2025enhancing} demonstrated that AST and Dataflow components most accurately capture structural similarity and program-level semantics of code. Therefore, we adopt their recommended weighting configuration ($\gamma = \delta = 0.5$) and proposed CodeBLEU formulation as follows:
\begin{equation}
\text{CodeBLEU}_{\text{reported}} =
\gamma \cdot \text{AST}
+ \delta \cdot \text{Dataflow}.
\end{equation}

\subsection{Empirical Evaluation}

\paragraph{Answer to RQ1.} Following SWEET and WLLM, we define watermarking effectiveness as the combined ability to (i) embed a statistically detectable signal in generated code and (ii) maintain high True Positive Rates (TPR) while keeping False Positive Rates (FPR) low. Table~\ref{tab:results} reports the full comparison across all baselines and proposed variants.

To systematically evaluate PromptMark, we structured the experiments in three stages. First, we compared our approach against two representative baselines: SWEET and In-Context Watermarking (ICW). Second, we conducted an ablation study of several prompt-driven watermarking variants to analyze the impact of different design choices. Finally, after identifying the most effective configuration, we evaluated the finalized framework on the corresponding datasets. The implementation details are available at \url{https://github.com/ahmedfahad04/promptmark}.

\textbf{Baseline Assessment.}
Among the baseline methods, the \textit{Initials In-Context Watermarking (ICW)} approach performs the worst, achieving an AUROC of only 0.51. Treating source code as plain natural text fails to bias identifier distributions sufficiently to produce a detectable watermark signal. In contrast, the SWEET watermark achieves higher detectability but significantly degrades functional correctness. This degradation occurs because SWEET perturbs token logits during decoding without enforcing syntax-aware constraints. Since code is structurally rigid, even small token-level perturbations can corrupt critical syntactic elements, thereby producing invalid code.

\textbf{Evaluation of Prompt-Based Variants.}
We next evaluated three prompt-driven watermarking variants to understand how prompt design influences watermark embedding.

The \textit{Static Watermarking} variant splits identifier initials into fixed green list and red list sets. This approach improves detectability (AUROC 0.92), indicating that identifier prefix constraints can bias the token distribution. However, the detection rate remains limited (T@1\%F = 0.39), suggesting insufficient watermark coverage.

The \textit{Refactoring-Based} variant attempts to embed watermarks by providing the solution code and asking the model to refactor it. However, this strategy performs worse than Static watermarking in terms of detectability. LLMs tend to preserve existing identifiers when provided with full source code. This strategy rarely performs any structural edits following specified constraints, which is consistent with prior findings that LLMs typically avoid modifying existing tokens unless strongly instructed~\cite{chen2025envisioning}. 

The \textit{Comment-Based} variant extends the watermarking channel by allowing comments to follow the green list constraint. This modification partially alleviates the limited identifier issue and improves detection performance, particularly at relaxed thresholds (T@5\%F = 0.68). However, the overall watermark signal remains lower than desired.

Taken together, these observations demonstrate that single-prompt watermarking strategies are insufficient for achieving both strong detectability and stable performance.

\textbf{Impact of Iterative Feedback.}
To address the limitations of single-pass prompting, an \textit{Iterative Watermarking} strategy is adopted, incorporating feedback during generation. Inspired by recent findings that iterative self-refinement improves instruction following behavior of LLMs~\cite{dong2025selfplay}, our framework repeatedly evaluates candidate outputs and provides corrective feedback to the model to self-evaluate its own response. This iterative mechanism substantially strengthens watermark detectability. On average, the model requires approximately 2 iterations to produce a valid watermarked program. The resulting outputs achieve significantly improved separability, with AUROC approaching 0.97 and higher detection rates under strict FPR constraints.

However, we noticed that the identifiers generated under strict green/red constraints often appear unnatural or forced, reducing code readability.

\textbf{Adaptive Green List Strategy (AGL).}
To address the naturalness issue, we introduced an \textit{Adaptive Green List} derived from the empirical identifier-initial distribution observed in human-written code (see Section~\ref{sub-freq-stratification}). By selecting watermark tokens from high-frequency initials, the model can generate identifiers that remain semantically meaningful and stylistically natural. Integrating this adaptive strategy with iterative watermarking (Iterative-AGL) produces the strongest overall detection performance. In particular, the Gemini-MBPP setting achieves the highest separability (AUROC = 0.994) with strong detection under strict constraints. 

While the Gemini-MBPP configuration achieves the highest scores, the other model-dataset combinations show only minor variations in performance, with AUROC values consistently remaining above 0.90 across all settings. In contrast, the baseline methods exhibit a larger performance drop under these settings, as reflected in Table~\ref{tab:results}. These variations mainly arise from differences in dataset characteristics and model-specific generation behavior. MBPP primarily consists of short utility-style programming tasks, whereas HumanEval focuses on more complex algorithmic problems. As a result, HumanEval solutions typically contain fewer user-defined identifiers and comments, which slightly reduces the watermark embedding capacity compared to MBPP. Despite these structural differences between the datasets, our proposed method consistently achieves strong watermark detectability across both benchmarks, supporting RQ1.

\paragraph{Answer to RQ2.} Across both datasets and models, PromptMark preserves functional correctness while embedding watermark signals. The correctness (Pass\%) remains consistently high across all configurations, scoring up to ~97\% on HumanEval and ~99\% on MBPP. This indicates that the watermarking constraints do not collide with program logic or execution behavior, and the generated programs remain functionally valid.

We measured the structural similarity among the generated code using CodeBLEU scores. Here, the Refactoring-Based variant produces the highest CodeBLEU scores overall, which are up to 0.91 for Gemini and 0.79 for Claude. This behavior is expected as the model receives the reference solution and tends to preserve the original structure rather than actively modifying identifiers to embed watermarks. As a result, the generated code remains highly similar to the ground-truth implementation, leading to inflated structural similarity while providing weaker watermark signals.

On the other hand, CodeBLEU remains stable across watermarking variants. Among the prompt-guided approaches, the Gemini-MBPP configuration achieves the highest CodeBLEU score (0.46), while Claude-HumanEval reaches a comparatively higher structural similarity (0.50), which closely aligns with the non-watermarked results (shown as the \textit{No Watermark} method). These results indicate that the watermarking process preserves control-flow structure, data dependencies, and syntactic organization of the generated code across different models and datasets.

Overall, our approach consistently outperforms baseline approaches while maintaining balanced functional correctness and structural integrity of the generated code, thereby supporting RQ2.

\section{\uppercase{DISCUSSION}}
In this section, we analyze PromptMark's capability for model-specific identification, its resilience to code transformations, and the trade-off between iteration cost and performance.

\subsection{Model-Specific Watermarking}

In our framework, the watermark is generated based on a secret key $k$. Let the green list be defined as $G = \text{GenGreen}(k)$, where $\text{GenGreen}(\cdot)$ denotes the deterministic cryptographic procedure described in Section~\ref{sub-freq-stratification}. If the same key $k$ is used across multiple models ($\mathcal{M}_1,\dots,\mathcal{M}_n$), then the green list $G$ will be the same. As a result, the watermark proves that the output was generated under key $k$, but it cannot identify which model produced it. In this case, attribution is tied to the key (i.e., the watermarking authority), not to the underlying model.

To enable model-specific attribution, each model $\mathcal{M}_i$ must be assigned a distinct secret key $k_i$, yielding $G_i = \text{GenGreen}(k_i)$. Since $k_i \neq k_j$ for $i \neq j$, the corresponding green-lists $G_i$ and $G_j$ are statistically independent. During detection, a given code sample $\hat{y}$ can be evaluated against multiple candidate keys $\{k_1, k_2, \dots, k_m\}$, producing detection scores $\{S_1, S_2, \dots, S_m\}$ (see Section ~\ref{subsec-wm-detection}). The attributed model index is then defined as $\hat{i} = \arg\max_i S_i$ provided that $S_i$ exceeds a predefined significance threshold. This approach follows the same principle as digital signatures: if two systems use the same key, they are indistinguishable, but if they use different keys, their outputs become uniquely identifiable~\cite{wu2024dipmark}. Thus, key separation enables model-specific attribution, while shared keys provide system-level attribution.

\begin{figure}
    \centering
    \includegraphics[width=0.99\linewidth]{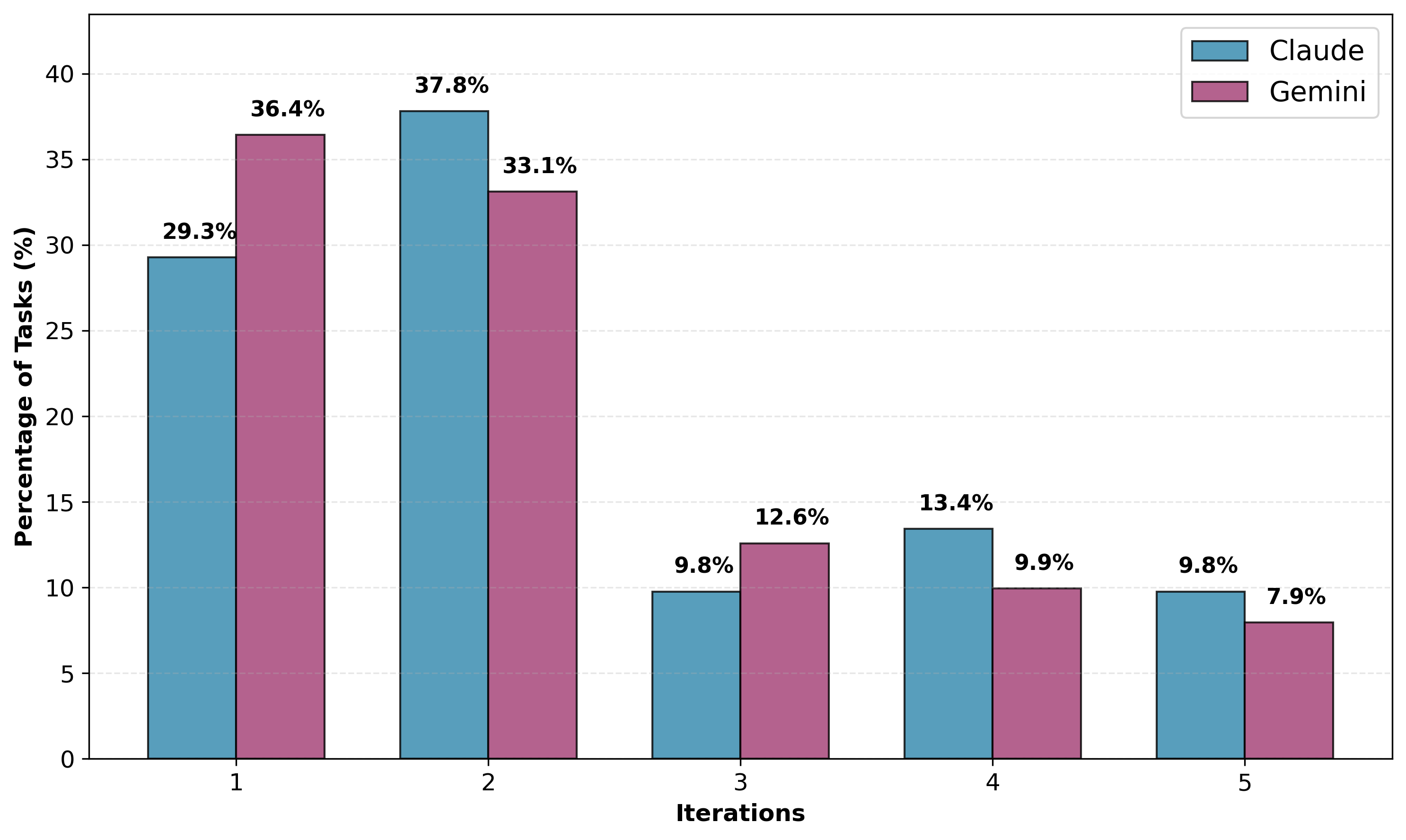}
    \caption{\centering{Iteration distribution across Claude and Gemini}}
    \label{fig:itr_dist}
\end{figure}

\begin{figure}
    \centering
    \includegraphics[width=0.99\linewidth]{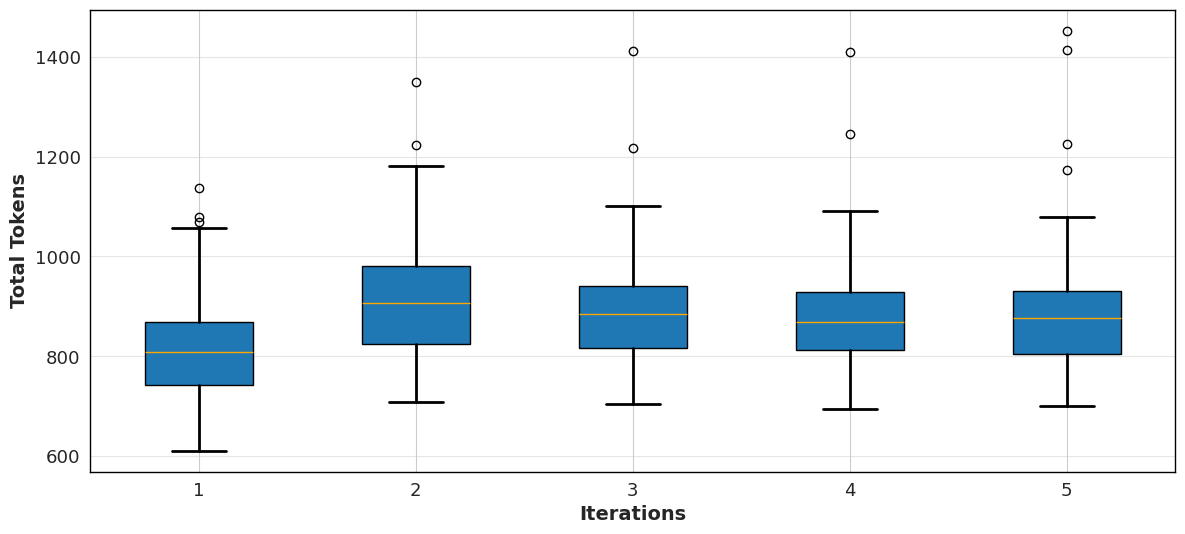}
    \caption{\centering{Box Plot of Total Token Count (Input and Output) Per Iteration}}
    \label{fig:token_dist}
\end{figure}

\subsection{Iteration Distribution and Cost Tradeoff}

Figure~\ref{fig:itr_dist} shows the iteration distribution for our generation procedure. Most tasks converge within the first two iterations, with only a small percentage requiring further refinement. This pattern aligns with prior work on iterative code refinement, such as AgentCoder ~\cite{huang2023agentcoder}, which also evaluates effectiveness across 1-5 iterations and finds incremental improvements with additional passes.

Figure~\ref{fig:token_dist} illustrates the total token usage aggregating input and output tokens per generation. It typically requires 800--900 tokens, which under standard pricing models of Claude Sonnet 4 translates into a negligible cost of approximately 1.4--1.6 cents per iteration. As cost scales with code length and complexity, and most tasks converge within a few iterations, the total expense remains minimal. Ultimately, this represents a pragmatic trade-off that the marginal computational cost is justified by the significant value of securing authorship attribution and verifiable code provenance.

\subsection{Robustness and Trade-Offs}

Watermark robustness under transformations is an important consideration, as prior works ~\cite{kim2025marking} have shown that watermarking schemes can degrade under paraphrasing or other semantic-preserving edits. To evaluate this aspect, we simulate two practical maintenance scenarios on the HumanEval dataset: (i) comment removal, and (ii) automated paraphrasing. We assume a black-box adversary who may have watermarked outputs and apply semantic-preserving transformations, but does not have access to the secret key. To imitate the attack scenario, we utilized GPT-Codex-5.3~\cite{openai_gpt53_codex_2026} with the specific instruction: \textit{``Paraphrase the scripts and rename the variables where necessary, keeping them relevant and natural.''}

\begin{table}[htbp]
    \centering
    \caption{\centering{Robustness of watermark detection under common maintenance attacks.}}
    \label{tab:robustness_attacks}
    \resizebox{\columnwidth}{!}{%
        \begin{tabular}{lccc}
            \toprule
            \textbf{Attack Type} & \textbf{AUROC} & \textbf{Degradation} & \textbf{Degradation \%} \\
            \midrule
            Baseline & 0.9834 & -- & -- \\
            Without Comments & 0.9456 & -0.0378 & -3.84\% \\
            Paraphrased & 0.7139 & -0.2695 & -27.41\% \\
            \bottomrule
        \end{tabular}%
    }
\end{table}

Table~\ref{tab:robustness_attacks} summarizes the detection performance under these transformations. On the original watermarked code, the detector achieves AUROC = 0.9834. After randomly stripping comments, the resulting AUROC decreases marginally to 0.9456. This suggests the watermark signal is distributed across identifiers rather than concentrated in comments. However, automated paraphrasing results in a substantial degradation, with AUROC dropping to 0.7139. 

These findings highlight an inherent trade-off among watermark strength, naturalness, and robustness. Stronger watermark signals, achieved through more pronounced lexical bias (e.g., enforcing green list initials), improve statistical detectability but may lead to less natural identifier naming. Conversely, aligning the watermark with natural identifier distributions improves readability but slightly limits signal strength. From a robustness perspective, lexical watermarking remains resilient to light maintenance (e.g., comment removal or minor edits) but is vulnerable to aggressive transformations such as systematic renaming or large-scale refactoring, which can disrupt the embedded statistical bias.  However, such attacks require substantial effort and changes, which may introduce new functional errors.
Techniques like multi-channel watermarking, such as combining semantic and syntactic channels, can enhance robustness for real-world deployments.





\section{\uppercase{Threats to Validity}}
\label{sec:threats}
The validity of our results depends on how PromptMark uses LLM instruction following behavior. This can cause variation in watermark strength as LLMs are inherently non-deterministic, producing inconsistent and different outputs on the same prompt. To reduce this effect, we run each experiment twice using two benchmark datasets and two different LLMs. Additionally, our iterative-feedback approach stabilizes prompt variation impacts, maintaining both correctness and watermark strength across generations.

To enable a controlled analysis, our empirical evaluation is currently focused on Python and two benchmark datasets as follows in prior studies \cite{zhang2025robust}. Consequently, the detection performance reported in this work is tied to Python's identifier distribution. However, languages with fundamentally different naming conventions may introduce challenges that are not observed in our Python evaluation. Therefore, future empirical studies are required to verify the cross-language effectiveness of our proposed approach.


\section{\uppercase{Conclusion and Future Work}}
\label{sec:conclusion}

This framework introduces a systematic approach that reframes black-box source code watermarking as a prompt-guided task. By shifting from intrusive logit manipulation to a statistically grounded, frequency-based stratification strategy, our approach cryptographically derives a key-dependent ``green list'' by analyzing human-authored code. This ensures the embedded signal is reproducible, secure, and linguistically natural. Coupled with an iterative feedback loop, the LLM dynamically balances watermark fidelity against rigid syntactic constraints. Experimental results demonstrated that prompting can serve as an effective approach to embed verifiable watermarks, preserving code functionality and structure. 

While this work establishes a foundation for zero-bit prompt-guided watermarking, it opens complex avenues for future research. A major challenge is developing \textit{encodable, multi-bit watermarking} exclusively for a black-box setting. Additionally, to tackle semantic-preserving adversarial attacks, future research should explore \textit{multi-channel watermark embedding} to achieve white-box-level robustness in black-box environments. 

\section*{\uppercase{Acknowledgments}}
The author(s) used ChatGPT for language refinement during manuscript preparation and take full responsibility for the final content. This research is funded by a fellowship from the ICT Division, Government of Bangladesh; No-56.00.0000.000.052.33.0001.25-159, dated 16.04.2025.

\bibliographystyle{apalike}
{\small
\bibliography{example}

@inproceedings{lee2024,
    title = "Who Wrote this Code? Watermarking for Code Generation",
    author = "Lee, Taehyun  and
      Hong, Seokhee  and
      Ahn, Jaewoo  and
      Hong, Ilgee  and
      Lee, Hwaran  and
      Yun, Sangdoo  and
      Shin, Jamin  and
      Kim, Gunhee",
    editor = "Ku, Lun-Wei  and
      Martins, Andre  and
      Srikumar, Vivek",
    booktitle = "Proceedings of the 62nd Annual Meeting of the Association for Computational Linguistics (Volume 1: Long Papers)",
    month = aug,
    year = "2024",
    address = "Bangkok, Thailand",
    publisher = "Association for Computational Linguistics",
    url = "https://aclanthology.org/2024.acl-long.268/",
    doi = "10.18653/v1/2024.acl-long.268",
    pages = "4890--4911",
}

@misc{murray_2025, 
    title={The Latest Cybersecurity Executive Order: Implications and Guidance}, 
    url={https://chertoffgroup.com/the-latest-cybersecurity-executive-order-implications-and-guidance/}, 
    journal={The Chertoff Group}, 
    author={Murray, Ellen}, 
    year={2025}, 
    month={Jun},
    note={Accessed: 2026-02-20},
}

@misc{gu_2025,
  author       = {Jing Gu},
  title        = {Why Is Code Provenance Non-Negotiable in the Age of AI?},
  year         = {2025},
  month        = {Dec},
  day          = {17},
  url          = {https://www.beyondidentity.com/resource/why-is-code-provenance-non-negotiable-in-the-age-of-ai},
  note         = {Accessed: 2026-02-20},
}

@article{xiang_review_25,
author = {Xiang, Lingyun and Li, Nian and Liu, Yuling and Hu, Jiayong},
year = {2025},
month = {01},
pages = {1-10},
title = {AI-Generated Text Detection: A Comprehensive Review of Active and Passive Approaches},
journal = {Computers, Materials \& Continua},
doi = {10.32604/cmc.2025.073347}
}

@article{xu2025large,
  title={Large Language Models Are Effective Code Watermarkers},
  author={Xu, Rui and Chen, Jiawei and Yin, Zhaoxia and Kong, Cong and Zhang, Xinpeng},
  journal={arXiv preprint arXiv:2510.11251},
  year={2025}
}

@inproceedings{gurioli2025you,
  title={Is This You, LLM? Recognizing AI-written Programs with Multilingual Code Stylometry},
  author={Gurioli, A. and Gabbrielli, M. and Zacchiroli, S.},
  booktitle={2025 IEEE International Conference on Software Analysis, Evolution and Reengineering (SANER)},
  pages={394--405},
  year={2025}
}

@article{yin2025detecting,
  title={Detecting LLM-generated Code with Subtle Modification by Adversarial Training},
  author={Yin, Xin and others},
  journal={arXiv preprint},
  year={2025}
}

@article{paek2025detection,
  title={Detection of LLM-Generated Java Code Using Discretized Nested Bigrams},
  author={Paek, Timothy and Mohan, C.},
  journal={arXiv preprint},
  year={2025}
}

@inproceedings{zhao2025sok,
  title={Sok: Watermarking for ai-generated content},
  author={Zhao, Xuandong and Gunn, Sam and Christ, Miranda and Fairoze, Jaiden and Fabrega, Andres and Carlini, Nicholas and Garg, Sanjam and Hong, Sanghyun and Nasr, Milad and Tramer, Florian and others},
  booktitle={2025 IEEE Symposium on Security and Privacy (SP)},
  pages={2621--2639},
  year={2025},
  organization={IEEE}
}

@article{xu2024distinguishing,
  title={Distinguishing LLM-Generated from Human-Written Code by Contrastive Learning},
  author={Xu, Xiaodan and others},
  journal={arXiv preprint},
  year={2024}
}

@article{park2025detecting,
  title={Detecting code paraphrased by large language models using coding style features},
  author={Park, Shinwoo and Jin, H. C. J-w. and others},
  journal={Engineering Applications of Artificial Intelligence},
  volume={162},
  pages={112454},
  year={2025}
}

@inproceedings{
    liu2025,
    title={In-Context Watermarks for Large Language Models},
    author={Yepeng Liu and Xuandong Zhao and Christopher Kruegel and Dawn Song and Yuheng Bu},
    booktitle={ICML 2025 Workshop on Reliable and Responsible Foundation Models},
    year={2025},
    url={https://openreview.net/forum?id=LE1kWLzeRs}
}

@article{li2024acw,
  title={ACW: Enhancing traceability of AI-generated codes based on watermarking},
  author={Li, Boquan and Zhang, Mengdi and Zhang, Peixin and Sun, Jun and Wang, Xingmei and Fu, Zirui},
  journal={arXiv preprint arXiv:2402.07518},
  year={2024}
}

@article{piantadosi2014zipf,
  title={Zipf’s word frequency law in natural language: A critical review and future directions},
  author={Piantadosi, Steven T},
  journal={Psychonomic bulletin \& review},
  volume={21},
  number={5},
  pages={1112--1130},
  year={2014},
  publisher={Springer}
}

@article{zhang2025robust,
  title={Robust and secure code watermarking for large language models via ml/crypto codesign},
  author={Zhang, Ruisi and Javidnia, Neusha and Sheybani, Nojan and Koushanfar, Farinaz},
  journal={arXiv preprint arXiv:2502.02068},
  year={2025}
}

@article{collberg_thomborson_2002, 
title={Watermarking, tamper-proofing, and obfuscation - tools for software protection}, volume={28}, 
url={https://ieeexplore.ieee.org/document/1027797}, 
DOI={https://doi.org/10.1109/tse.2002.1027797}, 
number={8}, journal={IEEE Transactions on Software Engineering}, 
publisher={Institute of Electrical and Electronics Engineers (IEEE)}, 
author={Collberg, C.S. and Thomborson, C.}, 
year={2002}, 
month={Aug}, 
pages={735–746} 
}

@article{hayes2017generating,
  title={Generating steganographic images via adversarial training},
  author={Hayes, Jamie and Danezis, George},
  journal={Advances in neural information processing systems},
  volume={30},
  year={2017}
}

@article{liu2024audiomarkbench,
  title={Audiomarkbench: Benchmarking robustness of audio watermarking},
  author={Liu, Hongbin and Guo, Moyang and Jiang, Zhengyuan and Wang, Lun and Gong, Neil},
  journal={Advances in Neural Information Processing Systems},
  volume={37},
  pages={52241--52265},
  year={2024}
}

@article{bistarelli_fiore_mercanti_mongiello_2025, title={Usage of Large Language Model for Code Generation Tasks: A Review}, volume={6}, url={https://link.springer.com/article/10.1007/s42979-025-04241-5#citeas}, DOI={https://doi.org/10.1007/s42979-025-04241-5}, number={6}, journal={SN Computer Science}, publisher={Springer Science and Business Media LLC}, author={Bistarelli, Stefano and Fiore, Marco and Mercanti, Ivan and Mongiello, Marina}, year={2025}, month={Jul} }

@article{kim2025marking,
  title={Marking Code Without Breaking It: Code Watermarking for Detecting LLM-Generated Code},
  author={Kim, Jungin and Park, Shinwoo and Han, Yo-Sub},
  journal={arXiv preprint arXiv:2502.18851},
  year={2025}
}

@article{austin2021programmbpp,
  title={Program Synthesis with Large Language Models},
  author={Austin, Jacob and Odena, Augustus and Nye, Maxwell and Bosma, Maarten and Michalewski, Henryk and Dohan, David and Jiang, Ellen and Cai, Carrie and Terry, Michael and Le, Quoc and others},
  journal={arXiv preprint arXiv:2108.07732},
  year={2021}
}

@misc{chen2021evaluatinghumaneval,
  title={Evaluating large language models trained on code},
  author={Chen, Mark and Tworek, Jerry and Jun, Heewoo and Yuan, Qiming and Pinto, Henrique Ponde De Oliveira and Kaplan, Jared and Edwards, Harri and Burda, Yuri and Joseph, Nicholas and Brockman, Greg and others},
  journal={arXiv preprint arXiv:2107.03374},
  year={2021}
}

@inproceedings{kirchenbauer2023watermark,
  title={A watermark for large language models},
  author={Kirchenbauer, John and Geiping, Jonas and Wen, Yuxin and Katz, Jonathan and Miers, Ian and Goldstein, Tom},
  booktitle={International Conference on Machine Learning},
  pages={17061--17084},
  year={2023},
  organization={PMLR}
}

@inproceedings{guan2024codeip,
    title = "{C}ode{IP}: A Grammar-Guided Multi-Bit Watermark for Large Language Models of Code",
    author = "Guan, Batu  and
      Wan, Yao  and
      Bi, Zhangqian  and
      Wang, Zheng  and
      Zhang, Hongyu  and
      Zhou, Pan  and
      Sun, Lichao",
    editor = "Al-Onaizan, Yaser  and
      Bansal, Mohit  and
      Chen, Yun-Nung",
    booktitle = "Findings of the Association for Computational Linguistics: EMNLP 2024",
    month = nov,
    year = "2024",
    address = "Miami, Florida, USA",
    publisher = "Association for Computational Linguistics",
    url = "https://aclanthology.org/2024.findings-emnlp.541/",
    doi = "10.18653/v1/2024.findings-emnlp.541",
    pages = "9243--9258"
}

@INPROCEEDINGS{srcmarker,
  author={Yang, Borui and Li, Wei and Xiang, Liyao and Li, Bo},
  booktitle={2024 IEEE Symposium on Security and Privacy (SP)}, 
  title={SrcMarker: Dual-Channel Source Code Watermarking via Scalable Code Transformations}, 
  year={2024},
  volume={},
  number={},
  pages={4088-4106},
  doi={10.1109/SP54263.2024.00097}
}

@article{li2025codeguard,
  title={CodeGuard: A Generalized and Stealthy Backdoor Watermarking for Generative Code Models},
  author={Li, Haoxuan and Zhang, Jiale and Sun, Xiaobing and Luo, Xiapu},
  journal={arXiv preprint arXiv:2506.20926},
  year={2025}
}

@article{dasgupta2024watermarking,
  title={Watermarking language models through language models},
  author={Dasgupta, Agnibh and Tanvir, Abdullah and Zhong, Xin},
  journal={arXiv preprint arXiv:2411.05091},
  year={2024}
}

@inproceedings{wu2024dipmark,
  title={DiPmark: A Stealthy, Efficient and Resilient Watermark for {LLM}s},
  author={Zhi-Yi Wu and Ling-Liang Wu and Jichuan Guan and Jihong Guan and Shuigeng Zhou},
  booktitle={International Conference on Learning Representations (ICLR)},
  year={2024},
  url={https://openreview.net}
}

@inproceedings{chen2025envisioning,
  title={Envisioning the Future of Peer Review: Investigating LLM-Assisted Reviewing Using ChatGPT as a Case Study},
  author={Chen, Shiping and Brumby, Duncan and Cox, Anna},
  booktitle={Proceedings of the 4th Annual Symposium on Human-Computer Interaction for Work},
  pages={1--18},
  year={2025}
}

@inproceedings{
dong2025selfplay,
title={Self-play with Execution Feedback: Improving Instruction-following Capabilities of Large Language Models},
author={Guanting Dong and Keming Lu and Chengpeng Li and Tingyu Xia and Bowen Yu and Chang Zhou and Jingren Zhou},
booktitle={The Thirteenth International Conference on Learning Representations},
year={2025},
url={https://openreview.net/forum?id=cRR0oDFEBC}
}

@article{li2011application,
  title={Application of volcano plots in analyses of mRNA differential expressions with microarrays},
  author={Li, Wentian},
  journal={arXiv preprint arXiv:1103.3434},
  year={2011}
}

@article{ning2024mcgmark,
  title={MCGMark: An Encodable and Robust Online Watermark for Tracing LLM-Generated Malicious Code},
  author={Ning, Kaiwen and Chen, Jiachi and Zhong, Qingyuan and Zhang, Tao and Wang, Yanlin and Li, Wei and Zhang, Jingwen and Yu, Jianxing and Feng, Yuming and Zhang, Weizhe and others},
  journal={arXiv preprint arXiv:2408.01354},
  year={2024}
}

@book{knuth1997art,
  author = {Knuth, Donald E.},
  title = {The Art of Computer Programming, Volume 2: Seminumerical Algorithms},
  edition = {3rd},
  year = {1997},
  publisher = {Addison-Wesley},
  address = {Reading, MA}
}

@online{openai_gpt53_codex_2026,
  author       = {{OpenAI}},
  title        = {{Introducing GPT-5.3-Codex}},
  year         = {2026},
  month        = feb,
  day          = {5},
  url          = {https://openai.com/index/introducing-gpt-5-3-codex/},
  note         = {Accessed: 2026-02-23},
}

@article{mahmud2025enhancing,
  title={Enhancing llm code generation with ensembles: A similarity-based selection approach},
  author={Mahmud, Tarek and Duan, Bin and Pasareanu, Corina and Yang, Guowei},
  journal={arXiv preprint arXiv:2503.15838},
  year={2025}
}

@article{huang2023agentcoder,
  title={Agentcoder: Multi-agent-based code generation with iterative testing and optimisation},
  author={Huang, Dong and Zhang, Jie M and Luck, Michael and Bu, Qingwen and Qing, Yuhao and Cui, Heming},
  journal={arXiv preprint arXiv:2312.13010},
  year={2023}
}
}

\end{document}